\documentclass[
    ,final            
  ]
  {aipproc}

\layoutstyle{8x11double}

\begin{document}

\title{Solitons in Supersymmetric Gauge Theories}


\author{M.~Eto}{
  address={Department of Physics, Tokyo Institute of 
Technology, Tokyo 152-8551, JAPAN}
}

\author{Y.~Isozumi}{
  address={Department of Physics, Tokyo Institute of 
Technology, Tokyo 152-8551, JAPAN}
}

\author{M.~Nitta}{
  address={Department of Physics, Tokyo Institute of 
Technology, Tokyo 152-8551, JAPAN}
}

\author{K.~Ohashi}{
  address={Department of Physics, Tokyo Institute of 
Technology, Tokyo 152-8551, JAPAN}
}

\author{N.~Sakai}{
  address={Department of Physics, Tokyo Institute of 
Technology, Tokyo 152-8551, JAPAN}
}

\begin{abstract}
Recent results on BPS solitons in the Higgs phase of 
supersymmetric (SUSY) gauge theories with eight 
supercharges are reviewed. 
For $U(N_{\rm C})$ gauge theories 
with the $N_{\rm F}(>N_{\rm C})$ hypermultiplets in the 
fundamental representation, 
the total moduli space of walls are found to be the complex 
Grassmann manifold 
$SU(N_{\rm F})/[SU(N_{\rm C})\times SU(N_{\rm F}-N_{\rm C})
\times U(1)]$. 
The monopole in the Higgs phase has to accompany 
vortices, and 
preserves a $1/4$ of SUSY. 
We find that walls are also 
allowed to coexist with them. 
We obtain all the solutions of such $1/4$ BPS composite 
solitons in the strong coupling limit. 
Instantons in the Higgs phase is also obtained as 1/4 BPS 
states. 
As another instructive example, we take $U(1)\times U(1)$ 
gauge theories with four hypermultiplets. 
We find that the moduli 
space is the union of several special 
Lagrangian submanifolds of the Higgs branch vacua 
of the corresponding massless theory. 
We also observe transmutation of 
walls and repulsion and attraction of BPS walls. 
This is a review of recent works on the subject, which was given 
at the conference by N.~Sakai. 
\end{abstract}

\maketitle


\section{
Higgs Phase Vacua 
and BPS Eq.
}

In recent years, models with extra dimensions are often 
used to obtain unified theories beyond the standard 
model \cite{HoravaWitten}. 
In this brane-world scenario, we need to construct a soliton 
whose world volume effective theory resembles the standard 
model. 
These solitons are usually some kind of topological defects 
of our higher dimensional theory. 
The simplest soliton is the domain wall with co-dimension one, 
and the next simplest is the vortex with co-dimension two, 
whereas the co-dimension three (four) soliton is called 
monopole (instanton). 
It has also been customary to consider supersymmetric 
(SUSY) theories in order to build realistic unified models 
\cite{DGSW}. 
Supersymmetric theories often help to obtain stable solitons 
as BPS states, which preserve part of supersymetry of the 
origonal theory \cite{WittenOlive}. 
It has been known that these BPS solitons automatically solve 
the field equations and their stability is usually guaranteed 
by topological charges. 
Moreover, these BPS solitons have been extremely useful to 
understand the nonperturbative dynamics of supersymmetric 
theories. 

In order to obtain low-energy effective theory on the world 
volume of the soliton, it is necessary to find massless 
modes, which are obtained by promoting the parameter of the 
soliton solution \cite{Ma}, namely moduli. 
In recent years, a subtancial progress has been made on 
understanding the moduli and their dynamics for supersymmetric 
gauge theories with eight supercharges, especially in its 
Higgs phase. 
The purpose of this paper is to report some of the progress. 

As a concrete model, we are primarily interested in SUSY 
$U(N_{\rm C})$ gauge theories with $N_{\rm F} > N_{\rm C}$ 
hypermultiplets in five or six dimensions. 
The minimal number of SUSY in these dimensions is eight. 
We can easily obtain theories in lower dimensions by making 
a simple or Scherk-Schwarz dimensional reduction. 
Let us first discuss domain walls. 
To obtain domain walls, we need to have two or more discrete 
vacua. 
It is only achieved by using massive hypermultiplets, which 
requires five or lower dimensions. 
The bosonic components of the vector multiplet in theories 
with eight SUSY consist of a gauge field $W_M$, and 
auxiliary fields $Y^a, a=1,2,3$, which are represented by 
$N_{\rm C}\times N_{\rm C}$ matrices. 
The bosonic components of the hypermultiplet contain 
a doublet of complex scalars $H^i, i=1,2$, which are 
denoted as $N_{\rm C}\times N_{\rm F}$ matrices. 
The $N_{\rm F}\times N_{\rm F}$ hypermutiplet 
mass matrix is denoted as $M$. 
With a common gauge coupling for $U(1)$ and $SU(N_{\rm C})$ 
gauge group, and a Fayet-Iliopoulos parameter $c$, 
the bosonic part of the Lagrangian in five dimensions is 
given by 
\begin{eqnarray}
&\!\!\!&\!\!\!{\cal L}|_{\rm boson}
= 
{\rm Tr}\biggl[-{1\over 2g^2} F_{MN}(W)F^{MN}(W)
+\frac{1}{g^2}({\cal D}_M \Sigma)^2 
\nonumber \\
&\!\!\!+&\!\!\!  {\cal D}^M H^i ({\cal D}_M H^i)^\dagger 
-(\Sigma H^i-H^iM)(\Sigma H^i-H^iM)^\dagger 
\nonumber \\
&\!\!\!+&\!\!\!
\frac{1}{g^2}\sum_{a=1}^3(Y^a)^2-cY^3  
+Y^a(\sigma^a)_{ij}H^j H^{i\dagger} 
\biggr]. 
\label{fundamental-Lag1}
\end{eqnarray}
Covariant derivatives are 
${\cal D}_M \Sigma = \partial_M \Sigma + i[ W_M , \Sigma ]$, 
${\cal D}_M H^{irA}
=(\partial_M \delta_s^r + i(W_M)^r{}_s)H^{isA}$, 
and the gauge field strength is 
$F_{MN}(W)=-i[{\cal D}_M , {\cal D}_N]
$. 
The indices $M, N=0, 1,\cdots, 4$ run over five-dimensions, 
and the mostly plus signature is used for the 
metric $\eta_{MN}={\rm diag}.(-1, +1, \cdots, +1)$. 
We assume non-degenerate mass 
and $m_A > m_{A+1}$ for all $A$.  

The SUSY vacua are specified by vanishing vacuum energy. 
Vanishing contribution from vector 
multiplet read 
\begin{eqnarray}
H^{1}  H^{1\dagger}  - H^{2} H^{2\dagger} 
=c\mathbf{1}_{N_{\rm C}},
\quad 
 H^2 H^{1\dagger}= 0.
\label{D-term-cond}
\end{eqnarray}
Vanishing contribution to vacuum energy 
from hypermultiplets 
gives 
\begin{eqnarray}
\Sigma H^i - H^i M
=0. 
\label{F-term-cond++}
\end{eqnarray}
The $U(N_{\rm C})$ gauge transformations allow us to 
choose the vector multiplet scalar to be diagonal 
$\Sigma ={\rm diag}(\Sigma_1, \Sigma_2, \cdots, 
\Sigma_{N_C})$. 
Eq.(\ref{F-term-cond++}) requires the 
vector multiplet scalar to be nonvanishing $\Sigma_s=m_A$ 
for those nonvanishing hypermultiplets $H^{irA}$. 
Since we assume non-degenerate masses for hypermultiplets, 
we find that only one flavor $A=A_r$ 
can be non-vanishing for each color component $r$ of 
hypermultiplet scalars $H^{irA}$ with \cite{ANS}, 
\cite{INOS1}, \cite{INOS2} 
\begin{eqnarray}
 H^{1rA}=\sqrt{c}\,\delta ^{A_r}{}_A,\quad H^{2rA}=0.
 \label{eq:hyper-vacuum}
\end{eqnarray}
Therefore vacua are characterized by choosing $N_{\rm C}$ 
labels $A_1, A_2, \cdots, A_{N_{\rm C}}$ 
out of $N_{\rm F}$ flavors, corresponding to the nonvanishing 
color components $H^{1rA}$. 
We shall denote this vacuum as 
$ \langle A_1 A_2 \cdots A_{N_{\rm C}}\rangle $. 
This vacuum is called color-flavor locked vacua. 
These discrete vacua allow domain walls which interpolate 
discretely different vacua at left $y=-\infty$ and at right 
infinity $y=\infty$. 
Therefore topological sectors for multi-walls are labeled 
by boundary conditions 
$ \langle B_1 B_2 \cdots B_{N_{\rm C}}\rangle $ 
at $y=-\infty$ 
and 
$ \langle A_1 A_2 \cdots A_{N_{\rm C}}\rangle $
at  $y=\infty$.
\begin{figure}[htb]
\includegraphics[width=7.8cm,clip]{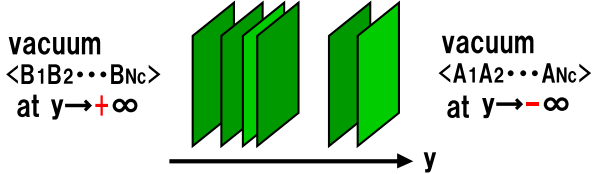}
\caption{
A multi-wall configuration connecting vacua 
$\langle A_1A_2\cdots A_{N_{\rm C}}\rangle$ 
and $\langle B_1B_2\cdots B_{N_{\rm C}}\rangle $. 
}
\label{su2nf}
\end{figure}

Let us consider co-dimension one soliton such as walls. 
We assume that all fields depend only on the coordinate 
of one extra dimension $x^4\equiv y$, and assume the 
Poincar\'e invariance on the four-dimensional world volume 
of the soliton. 
The well-known Bogomol'nyi completion of the energy 
density gives 
\begin{eqnarray}
{\cal E}
&\!\!\!\!\!=&\!\!\!\!\!
\frac{1}{g^2}{\rm Tr}\left({\cal D}_y \Sigma -
{g^2\over 2}\left(c{\bf 1}_{N_{\rm C}}-H^1H^1{}^\dagger 
+H^2H^2{}^\dagger \right)\right)^2
 \nonumber\\
&\!\!\!\!\!+&\!\!\!\!\!
{g^2}{\rm Tr}
\Big[
H^2H^{1\dagger} H^1H^{2\dagger} 
\Big] 
\nonumber \\
&\!\!\!\!\!+&\!\!\!\!\!
 {\rm Tr}\left[
({\cal D}_y H^1 + \Sigma H^1 -H^1M) 
({\cal D}_y H^1 + \Sigma H^1 -H^1M)^\dagger\right] 
\nonumber \\ &\!\!\!\!\!+&\!\!\!\!\!
 {\rm Tr}\left[
({\cal D}_y H^2 - \Sigma H^2 +H^2M) 
({\cal D}_y H^2 - \Sigma H^2 +H^2M)^\dagger\right]  
\nonumber \\ &\!\!\!\!\!+&\!\!\!\!\!
 c \partial_y{\rm Tr}\Sigma 
, 
\label{eq:bogomolnyi}
\end{eqnarray}
where we have omitted total divergence terms which give 
only vanishing contributions for topological charges. 
Thus we obtain the Bogomol'nyi bound as the lower bound 
for the energy of the soliton. 
By saturating the complete squares, 
we obtain the BPS equation 
\begin{eqnarray}
{\cal D}_y H^1 &\!\!\!=&\!\!\! -\Sigma H^1 + H^1 M,
\label{BPSeq-H1}
\\ 
{\cal D}_y H^2 &\!\!\!=&\!\!\! \Sigma H^2 -H^2 M,
\label{BPSeq-H2}
\\
{\cal D}_y \Sigma &\!\!\!=&\!\!\! 
{g^2\over 2}\left(c{\bf 1}_{N_{\rm C}}-H^1H^1{}^\dagger 
+H^2H^2{}^\dagger \right), 
\label{BPSeq-Sigma}
\\ 
0 &\!\!\!=&\!\!\! 
g^2 H^1H^2{}^\dagger . 
\end{eqnarray}
These conditions are precisely the condition for half of SUSY 
to be preserved by the soliton configuration. 
The energy (per unit world-volume of the wall) 
of the BPS saturated soliton for the topological sector 
labeled by 
$ \langle A_1 A_2 \cdots A_{N_{\rm C}}\rangle 
\leftarrow \langle B_1 B_2 \cdots B_{N_{\rm C}}\rangle $ 
is given by 
\begin{eqnarray}
T_{\rm w}=\int^{+\infty}_{-\infty}\hspace{-1.5em}dy{\cal E}
=c 
\left[{\rm Tr}\Sigma \right]^{+\infty}_{-\infty}
\!\!=\!c \left(\sum_{k=1}^{N_{\rm C}}m_{A_k}
-\sum_{k=1}^{N_{\rm C}}m_{B_k}\right).\qquad 
\label{eq:tension}
\end{eqnarray}

\section{BPS Wall Solutions}
\label{sc:bps-wall-solution}

\subsection{Solving BPS Equations}

Let us first 
introduce an 
$N_{\rm C}\times N_{\rm C}$ invertible 
complex matrix function $S(y)$ defined by 
\cite{INOS1}, \cite{INOS2} 
\begin{eqnarray}
\Sigma + iW_y \equiv S^{-1}\partial_y S.
\label{def-S}
\end{eqnarray}
By using this matrix function $S$ we obtain the solution 
of the hypermultiplet BPS equations (\ref{BPSeq-H1}) and 
(\ref{BPSeq-H2}) as 
\begin{eqnarray}
H^1=S^{-1}H_0 e^{My}, 
\quad H^2=0. 
\label{sol-H}
\end{eqnarray}
with the $N_{\rm C}\times N_{\rm F}$ constant 
complex matrices $H_0
$ as 
integration constants, which we call moduli matrices. 
We have used already the boundary condition for $H^2=0$ 
at $y=\pm \infty$.

Using the solution (\ref{sol-H}) 
of the hypermultiplet BPS equation, 
the remaining BPS equations (\ref{BPSeq-Sigma}) 
for the vector multiplets 
can be rewritten in terms of the matrix $S$ and the moduli 
matrix $H_0$. 
The $U(N_{\rm C})$ gauge transformations $U$ 
act on the matrix function $S$ as 
\begin{eqnarray}
 S\,\rightarrow \,S'=SU^\dagger ,\quad U^\dagger U=1. 
\label{eq:gauge-tr-S}
\end{eqnarray}
Thus we define a gauge invariant quantity $\Omega$ from 
$S$ as 
\begin{eqnarray}
 \Omega \equiv SS^\dagger.  
\label{def-Omega}
\end{eqnarray}
The moduli matrix $H_0$ is also gauge invariant. 
The BPS equations (\ref{BPSeq-Sigma}) for vector 
multiplets can be rewritten in the following 
gauge invariant form 
\begin{eqnarray}
 \partial _y\left(\Omega ^{-1}\partial _y\Omega \right)
=g^2c\left({\bf 1}_{\rm C}-\Omega ^{-1}\Omega _0\right),
\label{eq:master-eq-wall}
\end{eqnarray}
where the source term is given in terms of the moduli 
matrix $H_0$ as 
\begin{eqnarray}
\Omega _0\equiv c^{-1}H_0e^{2My}H_0^\dagger. 
\end{eqnarray}
We call Eq.(\ref{eq:master-eq-wall}) as 
the master equation for domain walls.

We should solve the master equation for a given moduli 
matrix $H_0$. 
It has been conjectured that the solution 
of the master equation always exists 
and is unique, for any given moduli 
matrix $H_0$ \cite{INOS1}, \cite{INOS2}. 
If this is true, the moduli matrix $H_0$ is the 
necessary and sufficient data for the moduli of 
the solution. 
This conjecture has been proved for $U(1)$ gauge thoeries 
recently \cite{SakaiYang}. 
For non-Abelian gauge theories such as $U(N_{\rm C})$, 
the best evidence for the moduli matrix to be the necessary 
and sufficient data for the moduli, is given by the index 
theorem \cite{Sakai:2005sp} : 
the number of independent 
parameters contained in the moduli matrix agrees precisely 
with that required by the index theorem. 
Additional evidence is provided by the exact solutions 
at strong gauge coupling limit and at discrete finite 
coupling\cite{IOS},  where the solution of master equation indeed 
exists for a restricted class of moduli matrices 
\cite{INOS1}, \cite{INOS2}.

\subsection{Total Moduli Space}

The differential equation (\ref{def-S}) defines 
the matrix function $S(y)$ only up 
$N_{\rm C}^2$ arbitrary complex integration constants. 
Therefore a set 
$(S, H_0)$ of the matrix function $S$ and the moduli 
matrix $H_0$ and 
another set $(S', H_0{}')$ give 
the same physical fields $\Sigma ,\,W_y, H^i$, 
provided they are related by 
\begin{eqnarray}
S' = VS,\quad 
H_0{}'=VH_0, 
\label{art-sym}
\end{eqnarray} 
where  
$V\in GL (N_{\rm C},{\bf C})$. 
We call this symmetry as `world-volume symmetry'. 
This equivalence relation $(S, H_0) \sim (S', H_0{}')$ 
defines an equivalence class of moduli matrices, 
which is the genuinely independent moduli. 
Namely the moduli space for (multi-)wall solutions 
(including all possible boundary conditions)
denoted by ${\cal M}_{N_{\rm F},N_{\rm C}}$
is topologically isomorphic to 
 the complex Grassmann manifold:
\begin{eqnarray}
 {\cal M}_{N_{\rm F},N_{\rm C}}
 &\!\!\!=
 &\!\!\!
\{H_0 | H_0 \sim V H_0, V \in GL(N_{\rm C},\mathbf{C})\} 
\nonumber\\
\equiv 
G_{N_{\rm F},N_{\rm C}}
 &\!\!\!\simeq &\!\!\! {SU(N_{\rm F}) \over 
 SU(N_{\rm C}) \times SU(N_{\rm F}-N_{\rm C}) 
 \times U(1)}\,,\qquad 
  \label{Gr}
\end{eqnarray} 
whose complex dimension is given by 
$N_{\rm C} (N_{\rm F}-N_{\rm C})$. 
This is a {\it compact} (closed) set. 
On the other hand, one expects noncompact moduli parameters, 
such as positions of walls. 
The presence of noncompact moduli parameters and the 
compactness of the total moduli space can be consistently 
understood, if we note that the moduli space 
${\cal M}_{N_{\rm F},N_{\rm C}}$ includes all topological sectors 
determined by different boundary conditions. 

\begin{figure}
\includegraphics[width=6cm,clip]{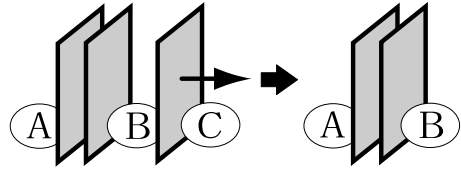}
\caption{
A three wall configuration connecting vacuum A to C 
through B (left). 
By letting the left-most wall to infinity, we obtain a 
two wall configuration connecting vacuum A to B. 
}\label{nit-fig2}
\end{figure}
Suppose we have a three wall configuration connecting 
the vacuum A at left infinity to the vacuum C at right 
infinity through the next to right vacuum B. 
If we let the right-most wall to the right infinity, 
we eventually obtain a two wall configuration connecting 
the vacuum A to the vacuum B. 
In this way, we can obtain two wall toplogical sectors as 
boundaries of a three wall topological sector. 
From the above illusrative example, we can observe 
that the Grassmann manifold as the total moduli space 
can be decomposed into various topological sectors 
${\cal M}^k$ for $k$ BPS walls 
\begin{eqnarray}
 {\cal M}_{N_{\rm F},N_{\rm C}} 
&\!\!\!\!
 = 
&\!\!\!\!
{\cal M}^{1/1} 
 \oplus {\cal M}^{1/2}, \quad 
 {\cal M}^{1/1}  = {\cal M}^0, 
\nonumber \\
{\cal M}^{1/2}
&\!\!\!\!
= 
&\!\!\!\!
{\cal M}^1
 \oplus \cdots 
  \oplus {\cal M}^{N_{\rm C}(N_{\rm F}-N_{\rm C})}. 
\end{eqnarray}
One should note that the total moduli space 
also includes the vacuum states with no walls 
$\langle A_1A_2\cdots A_{N_{\rm C}} \rangle$
$\leftarrow$ $\langle A_1A_2 \cdots A_{N_{\rm C}} \rangle$ 
which correspond to 
${}_{N_{\rm F}}C_{N_{\rm C}}$ points 
${\cal M}^0$. 
These vacua are the only points where all SUSY are preserved 
${\cal M}^{1/1}$. 
We can decompose in more detail according to topological 
sectors with specific boundary conditions 
\begin{eqnarray}
 {\cal M}_{N_{\rm F},N_{\rm C}} 
 = \sum_{\rm BPS} 
 {\cal M}^{\langle A_1A_2\cdots A_{N_{\rm C}}\rangle 
 \leftarrow \langle B_1B_2\cdots 
 B_{N_{\rm C}}\rangle }, 
\end{eqnarray}
where ${\cal M}^{\langle A_1A_2\cdots A_{N_{\rm C}}\rangle 
\leftarrow  
\langle B_1B_2 \cdots 
B_{N_{\rm C}}\rangle }
$
denotes the moduli subspace 
of BPS (multi-)wall solutions for the topological sector of 
$\langle A_1A_2\cdots A_{N_{\rm C}}\rangle$
$\leftarrow \langle B_1B_2 \cdots B_{N_{\rm C}}\rangle $, 
and the sum is taken over the BPS sectors. 
Although each sector (except for vacuum states) is in 
general an open set containing noncompact moduli, 
the total space is compact.

\subsection{Effective Lagrangian on BPS Walls}

In order to obtain the low-energy effective field theory, 
we need to promote the moduli parameters to fields 
$\phi^i(x), \phi^{i*}(x)$ on the world-volume of walls 
\cite{Ma}. 
It is again useful to use the solution (\ref{sol-H}) 
of the hypermultiplet BPS equation (\ref{BPSeq-H1}) 
in terms of the matrix function $S$. 
By using the solution of the hypermultiplet BPS equation, 
we can rewrite the Lagrangian in terms of the gauge 
invariant matrix $\Omega$. 
By systematically expanding the Lagrangian in powers 
of the slow-movement parameter $\lambda$, we can retain up 
to two powers of $\lambda$. 
We find the resulting effective Lagrangian as 
\begin{eqnarray}
{\cal L}
=
-T_{\rm w}+\int d ^4\theta K(\phi ,\phi ^*)
+{\rm higher}\; {\rm derivatives}, 
\end{eqnarray}
where $T_{\rm w}$ is the tension of the BPS (multi-)wall 
in Eq.(\ref{eq:tension}), and 
$K$ is the K\"ahler potential of moduli fields $\phi$ and 
$\phi^*$ 
\begin{eqnarray}
K(\phi ,\phi ^*)
&\!\!\!=&\!\!\!
\int dy \Bigl[
c\log{\rm det}\Omega 
+c{\rm Tr}\left(\Omega _0\Omega ^{-1}\right) 
\nonumber \\
&\!\!\!+&\!\!\!{1\over 2g^2}{\rm Tr}
\left({\Omega }^{-1}\partial _y\Omega \right)^2 
\Bigr]
\Big|_{\Omega =\Omega _{\rm sol}. 
}
\label{eq:kahler-pot}
\end{eqnarray}
We should replace the $\Omega$ in (\ref{eq:kahler-pot}) 
by the solution $\Omega_{\rm sol}$ of the master 
equation (\ref{eq:master-eq-wall}). 

It is interesting to observe that this K\"ahler potential 
serves as an action from which one can derive the master 
equation (\ref{eq:master-eq-wall}) by the usual 
minimal action principle. 
This fact can be explained if we use the superfield 
formulation for preserved four SUSY. 
By expanding the fundamental Lagrangian in powers of the 
slow-movement parameter $\lambda$, we find that the second 
hypermultiplet superfield $H^2$ and vector superfield $V$ 
become Lagrange multipliers. 
One can use the constraint equation resulting from 
integrating $H^2$ to obtain a Lagrangian in 
five dimensions amounting the Lagrangian for the remaining 
degree of freedom expressed in terms of the superfield 
$\Omega$. 
This is precisely the density of the K\"ahler potential 
in Eq.(\ref{eq:kahler-pot}) before integrating over $y$.

\subsection{Exact Solution at Strong Coupling}

In the strong gauge coupling limit $g^2\rightarrow \infty$, 
the master equation can be algebraically solved 
\begin{eqnarray}
 \Omega
 = \Omega_0 \equiv 
 c^{-1}H_0 e^{2My}H_0^\dagger. 
\label{eq:strong-coup}
\end{eqnarray}
With this solution, we can obtain $S$ by fixing a gauge. 
Then all the other fields such as $H^1$ and $\Sigma$ 
can also be obtained. 
There are two dimensionful parameters in the system: 
mass difference $\Delta m$ of hypermultiplets, and 
the gauge coupling times the square root of the 
Fayet-Iliopoulos parameter $g\sqrt{c}$. 
The strong coupling limit actually implies the limit 
$g^2c/\Delta m \gg 1$. 
This exact solution of master equation 
at $g^2\rightarrow\infty$ corresponds to 
the gauge theory becoming a nonlinear sigma model 
(NLSM) whose target space is the cotangent bundle over 
the Grassmann manifold $T^*G_{N_{\rm F},N_{\rm C}}$ 
\cite{ANS}. 
Since the Grassmann manifold is symmetric under the exchange 
of $N_{\rm C}$ and 
$\tilde N_{\rm C}\equiv N_{\rm F}-N_{\rm C}$ with fixed 
$N_{\rm F}$, 
all the result are also symmetric under the exchange 
(actually the sign of the Fayet-Iliopoulos parameter should 
be changed). 
We call this as a duality. 

Domain walls in this model can be realized 
as 
kinky D$p$-brane connecting separated D($p+4$)-branes 
in the type IIA/IIB string theory \cite{EINOOS}. 
Dynamics of domain walls can be understood very easily 
by this brane configuration.

\section{Global Structure of Wall Moduli Space}

We have also studied slightly different model where 
several new features of walls are realized. 
This is a $U(1)\times U(1)$ gauge theory with 
4 hypermultiplets $H^{1A}, H^{2A} \; (A=1,\cdots,4)$ 
with unequal charges \cite{Eto:2005wf}. 
The $U(1)\times U(1)$ charges for these 4 hypermultiplets 
are given by 
\begin{eqnarray}
\left( 
\begin{array}{cccc}
1& 1& 0& 0\\
0&-n& 1& 1\\
\end{array} 
\right). 
\end{eqnarray}
In the strong coupling limit $g^2\rightarrow \infty$, 
two $U(1)$ vector multiplets produces two constraints, 
and gives a NLSM. 

In order to study the constraints and the BPS flows for 
the multi-wall 
configurations, we define the following quantity 
as bilinears of scalar fields of 
the $A$-th hypermultiplet  
\begin{eqnarray}
\mu_A\equiv H^{1A\dagger}H^{1A} -H^{2A\dagger}H^{2A}, 
\quad 
\nu_A \equiv H^{2A\dagger} H^{1A} .
\end{eqnarray}
The conditions for SUSY vacua are given by 
\begin{eqnarray}
 \mu_1 + \mu_2
=
c_1, 
\qquad
 -n\mu_2 + \mu_3 + \mu_4
=
c_2, 
\label{eq:vacuum-cod1}
\end{eqnarray}
\begin{eqnarray}
 \nu_1 + \nu_2
=
0, 
 \qquad -n\nu_2 + \nu_3 + \nu_4
=
0. 
\label{eq:vacuum-cod2}
\end{eqnarray}

By studying the second constraint (\ref{eq:vacuum-cod2}), 
we find that it is not allowed to have nonvanishing values 
of both $H^{1A}$ (the first hypermultiplet) and $H^{2A}$ 
(the second hypermultiplet) for each flavor. 
This implies that the BPS flows occur only in a submanifold 
with half of the total dimensions. 
We find that this moduli space of the BPS flows is precisely 
a special Lagrangian submanifold of the target space of the 
NLSM. 
We find that there are several special Lagrangian 
submanifolds for this target space of the NLSM. 
The total moduli space is obtained as the union of all 
these special Lagrangian 
submanifolds. 

\begin{figure}
\includegraphics[width=13cm,clip]{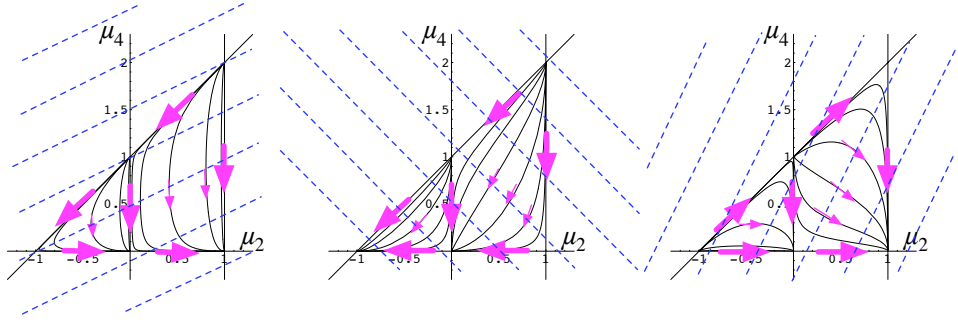}
\caption{
BPS flow in $n=1$ case.
For all cases $c_I=(1,1)$. 
{}From left to the right: case I) $m^A=(0,0,1,-1)$ ;
case II) $m^A=(1,0,0,-1)$ ;
case III) $m^A=(-1,0,1,0)$. 
Dashed lines designate the contours of constant $m^A\mu_A$.
}
\label{numericalflow}
\end{figure}

\begin{figure}
\includegraphics[width=.4\textwidth]{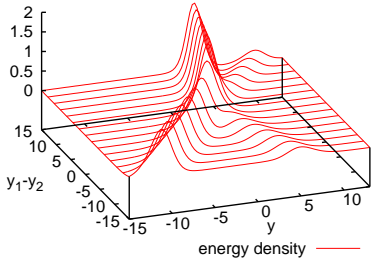}
\caption{
Transmutation of walls when they pass through.
$m^A=(1,0,0,-1)$ case. 
\label{exchange1}}
\end{figure}

\begin{figure}
\includegraphics[width=.4\textwidth]{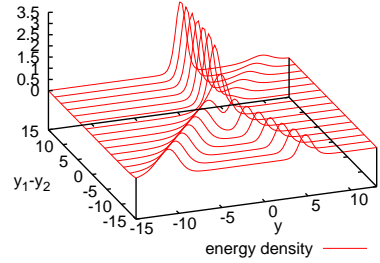}
\caption{
Transmutation of walls when they pass through.
$m^A=(2,0,0,-1)$ case.
\label{exchange2}}
\end{figure}

Because of two constraints from two $U(1)$ gauge groups, 
we only need to use two out of four 
$\mu_A, A=1, \cdots, 4$ to describe the BPS flows. 
Since the BPS flows stop only at vacua, 
the first vacuum conditions (\ref{eq:vacuum-cod1}) 
constitute boundaries of the BPS flows. 
Then the BPS flows for multi-wall configurations 
are restricted to closed polygons in the $\mu_1, \mu_2$ 
plane. 
We have shown three representative BPS flows depending on 
the mass assignments in Fig.\ref{numericalflow}. 
The case I) in Fig.\ref{numericalflow} shows that 
the three wall configuration has only two position moduli. 
This implies that there are repulsion 
and attraction between BPS walls, resulting in the 
phenomenon that the middle wall position of three 
walls is fixed. 
The case III) in Fig.\ref{numericalflow} shows that 
there is a transmutation of walls when two walls 
collide through moving in the moduli space. 
We have explicitly illustrated this movement in 
Fig.\ref{exchange1} and \ref{exchange2}. 

We have also observed that the 
moduli space dimension 
can be larger than 
naively suggested by index theorem 
for walls connecting a particular set of vacua, 
for instance two vacua on $\mu_2=0$ 
in the case I.

\section{Composite Solitons of Wall, Vortex and Monopole}

\subsection{Vortex Can Stretch between Walls}

\begin{figure}
\includegraphics[width=5.5cm,clip]{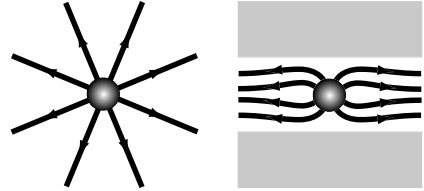}
\caption{
Magnetic flux from a monopole is spread in unbroken phase 
(left), but is squeezed in the Higgs phase (right). 
\label{fig:mihp}}
\end{figure}

So far we have been considering walls which can appear in 
the Higgs phase of SUSY gauge theories. 
We can have solitons with more codimensions, such as 
vortices, monopoles, and instantons. 
If $U(1)$ gauge group is unbroken, magnetic flux from a 
monopole can spread radially. 
However, if it is placed in the Higgs phase where there is no 
unbroken subgroups, magnetic flux has to be squeezed into a 
flux tube because of the Meissner effect, as illustrated in 
Fig.\ref{fig:mihp}. 
This composite soliton of monopoles and vortices has been 
found recently \cite{Tong:2003pz}, \cite{Hanany:2004ea}. 

Therefore a monopole in the Higgs phase has to accompany 
vortices. 
Although both monopoles and vortices are $1/2$ BPS solitons, 
they preserve different halves of SUSY. 
We find that they preserve only a quarter of 
SUSY, and that walls are also allowed at the same time 
as composite soliton preserving $1/4$ of SUSY \cite{Isozumi:2004vg}.  

We assume that the 
configuration depends on $x^m\equiv (x^1, x^2, x^3)$ 
(co-dimension three), and the Poincar\'e invariance 
in $x^0,x^4$. 
By requring a $1/4$ SUSY to be conserved, we obtain the 
$1/4$ BPS equations for the composite solitons 
\begin{eqnarray}
{\cal D}_3 \Sigma
&\!\!\!\!
=
&\!\!\!\!
{g^2 \over 2}
\left(c{\bf 1}_{N_{\rm C}}-H^1H^1{}^\dagger\right)
+F_{12} ,
\label{eq:wvmBPS1}
\\
{\cal D}_3 H^1 
&\!\!\!\!
=
&\!\!\!\!
-\Sigma H^1 +H M, \quad 
0={\cal D}_1 H^1 +i{\cal D}_2 H^1,
\label{eq:wvmBPS2}
\\
0
&\!\!\!\!
=
&\!\!\!\!
F_{23}
-{\cal D}_1 \Sigma, \quad 
0=F_{31}
-{\cal D}_2 
\Sigma, 
\label{eq:wvmBPS3}
\end{eqnarray}
where a contribution of vortex magnetic field $F_{12}$ 
is added to the wall BPS Eqs.(\ref{BPSeq-H1}), 
(\ref{BPSeq-Sigma}) together with 
the BPS equations for vortices. 
We obtain the BPS bound of the energy density as 
\begin{eqnarray}
&\!\!\!&\!\!\!{\cal E}\geq t_{\rm w}+t_{\rm v}+t_{\rm m}+\partial _mJ_m, 
\\
&\!\!\!&\!\!\! 
t_{\rm w}=c\partial _3 {\rm Tr}(\Sigma), \quad 
 t_{\rm v}=-c{\rm Tr} ( F_{12}),  
\nonumber \\ 
&\!\!\!&\!\!\!
 t_{\rm m}=
\frac{2}{g^2}\partial _m {\rm Tr}( \frac12 \epsilon^{mnl} \! 
F_{nl}\Sigma ) ,
\nonumber
\end{eqnarray}
where 
$t_{\rm w},\,t_{\rm v}$ and $t_{\rm m}$ are 
energy densities 
for walls, vortices and monopoles.

\subsection{Solutions of $1/4$ BPS Equations}

Eq.(\ref{eq:wvmBPS3}) guarantees the 
integrability condition of Eq.(\ref{eq:wvmBPS2}) 
\begin{eqnarray}
[{\cal D}_1+i{\cal D}_2,\,{\cal D}_3+\Sigma]=0, 
\end{eqnarray}
which allows us to define an invertible complex 
matrix function $S(x^m)\in GL(N_{\rm C},\mathbf{C})$ 
\begin{eqnarray}
({\cal D}_3+\Sigma)S^{-1}=0 
&\!\!\!
\rightarrow 
&\!\!\!
\Sigma + iW_3 \equiv S^{-1}\partial_3 S, \\
({\cal D}_1+i{\cal D}_2)S^{-1}=0 
&\!\!\!
\rightarrow 
&\!\!\!
W_1+iW_2\equiv -2iS^{-1}\bar \partial S, 
\nonumber
\end{eqnarray}
where $z \equiv x^1 + i x^2$, and 
$\bar \partial\equiv \partial/\partial z^*$. 
We can solve the hypermultiplet 
BPS Eq.(\ref{eq:wvmBPS2}) by 
\begin{eqnarray}
H^1 = S^{-1}(z,z^*, x^3)H_0(z) e^{M x^3}, 
\label{sol-H-vortex}
\end{eqnarray}
with the moduli matrix 
$H_0(z)$ as an $N_{\rm C} \times N_{\rm F}$ 
matrix as a holomorphic functionof $z$. 
The remaining BPS equation for the vector multiplet scalar 
can be recast into the master equation for a 
$U(N_{\rm C})$-gauge invariant matrix function 
$\Omega \equiv SS^\dagger $ 
\begin{eqnarray}
4\partial(\Omega ^{-1} \bar \partial \Omega) 
+\partial _3(\Omega ^{-1} \partial _3\Omega )
= g^2 
\left(c - \Omega ^{-1}\Omega_0
\right), 
\label{eq:master-eq-wvm}
\end{eqnarray}
with $\Omega_0
\equiv H_0 \,e^{2My} H_0{}^\dagger$ as a source term. 
This is an evolution equation along $x^3, z^*$, 
with the initial data $H_0(z)$, giving 
all possible solutions of the $1/4$ BPS equation 
in the $3$-dimensional 
configuration space. 
From the solution of the master equation for a given 
moduli matrix $H_0(z)$, we can obtain 
$S$ by fixing gauge, and then we also obtain 
$\Sigma ,\,W_m$ and $H^1$. 

Assuming existence of unique solution of this equation 
for $\Omega$, the moduli matrix $H_0(z)$ should 
contain the complete moduli of $1/4$ BPS soliton. 
Similarly to the wall case, we have also the world-volume 
symmetry 
\begin{eqnarray}
 H_0 \rightarrow H_0' = V H_0, \quad
 S \rightarrow S' = V S  \label{world-volume-tr}
\end{eqnarray}
with $V(z)$ an element of $GL(N_{\rm C},{\bf C})$ 
whose components are holomorphic in $z$. 
Then the total moduli space ${\cal M}_{\rm wvm}$ 
including {\it all} topological sectors with 
different boundary conditions 
can be identified as a quotient 
of the holomorphic maps defined by
\begin{eqnarray}
&&{\cal M}_{\rm wvm} =   
{\cal H}\backslash {\cal G}, 
\label{eq:moduli-space}
\\ 
&&{\cal G}  \equiv  
\{H_0 \ |\ {\bf C}^2 
{\longrightarrow} 
M(N_{\rm C} \times N_{\rm F}, {\bf C}), 
\bar\partial H_0 =0\}, 
\nonumber \\ 
&&{\cal H}  \equiv  \{V \ |\  {\bf C}^2 
{\longrightarrow} GL(N_{\rm C},{\bf C}),
\bar\partial V =0\},
\nonumber 
\end{eqnarray} 
where $M(N_{\rm C} \times N_{\rm F}, {\bf C})$ 
is an $N_{\rm C} \times N_{\rm F}$ complex matrix.
In the strong coupling limit, the moduli space reduces to 
the space of all the holomorphic maps from the complex $z$ plane 
to the complex Grassmann manifold 
\begin{equation}
 {\cal M}_{\rm wvm}^{g^2 \rightarrow\infty} 
 = \{\varphi| {\bf C}^2 \rightarrow G_{N_{\rm F},N_{\rm C}}, 
   \bar \partial_z \varphi  = 0 \}.
  \label{wvm-moduli}
\end{equation}

\subsection{Exact Solutions at Strong Coupling}

If we take the strong coupling limit $g^2\rightarrow \infty$, 
the master equation (\ref{eq:master-eq-wvm}) reduces to an 
algebraic equation 
\begin{eqnarray}
 \Omega
= \Omega_{0} 
\equiv c^{-1}H_0 e^{2My}H_0^\dagger . 
\end{eqnarray}
In this case, we can construct all solutions of 
the 1/4 BPS equations (\ref{eq:wvmBPS1})-(\ref{eq:wvmBPS3}) 
exactly and explicitly.

Our construction produces rich contents, even for 
the $U(1)$ gauge theories ($N_{\rm C}=1$). 
A general parametrization of the moduli matrix in this case 
is given by 
\begin{eqnarray}
H_0(z)=\sqrt{c}\left(f^1(z),\,\dots,f^{N_{\rm F}}(z)\right). 
\end{eqnarray}
In the strong coupling limit, the model reduces to 
a massive $T^*{\bf C}P^{N_{\rm F}-1}$ NLSM, where massive 
implies the presence of potential terms coming from the 
mass differences of hypermultipelts. 
The gauge invariant quantity is 
given in this $U(1)$ case by 
\begin{eqnarray}
\Omega =\sum_{A=1}^{N_{\rm F}}|f^A(z)|^2e^{2m_Ax^3}. 
\end{eqnarray}
Since each term in the sum of $\Omega$ corresponds to the 
weight of the vacuum, this can be regarded as a $z$ dependent 
multi-wall configuration. 
For each fixed $z$, we have maximally $N_{\rm F}-1$ walls 
at various points in $x^3$. 
The position of the $A$-th wall is given by 
$x_A^3(z)={\log|f_{A+1}(z)|-\log|f_A(z)|\over m_A-m_{A+1}}$. 
We now see that walls are bent for nonconstant $f^A(z)$. 
In particular, if $f^A(z)$ has 
zeroes, they should correspond to walls extending to infinity: 
namely vortices are formed. 
More precisely we find that 
$f^A(z) \propto (z-z^A_{\alpha })^{k^A_{\alpha }}$ 
produces a configuration with 
vorticity $k_{\alpha}^A$ 
at $z=z^A_{\alpha}$ on the $A$-th wall. 

\begin{figure}[htb]
\includegraphics[width=6cm, clip]
{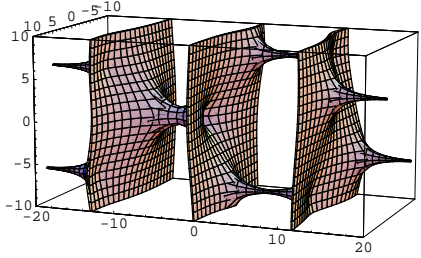}
\caption{\label{fig:wormhole} 
Vortices streched between multi-walls. 
Surfaces defined by the same energy density 
 with 
$t_{\rm w}+t_{\rm v}=0.5c$. 
}
\end{figure}
A typical $1/4$ composite soliton solution is depicted in 
Fig.\ref{fig:wormhole}. 
We note that the monopole in Higgs phase 
is realized as a kink on vortex world volume. 
Let us emphasize that our method allows complete solutions 
of vortex stretching between two or more walls. 
Thus generalizing the D-brane soliton as a BIon on a single 
wall (D-brane).

More recently we have also constructed a similar 
composite soliton with codimension four. 
The instantons in the Higgs phase should also accompany 
voritces in the Higgs phase. 
Following the suggestion in Ref.\cite{Hanany:2004ea}, 
we have obtained another $1/4$ BPS equation for 
the instantons in the Higgs phase, and have 
constructed a $1/4$ BPS solution \cite{EINOS}. 
Moreover, we also observed that the monopole in the Higgs 
phase can be obtained by a Scherk-Schwarz dimensional 
reduction from the instantons in the Higgs phase. 
To illustrate the mechanism more explicitly, it is useful 
to consider the calorons in the Higgs phase, 
which are the periodic array of 
instantons in the Higgs phase. 
Although the solution can be understood physically 
as a semi-local vortex (lump) on a vortex, 
our solution is a genuine solution of the $1/4$ BPS equation, 
and not just a $1/2$ BPS solution of the effective theory 
on the vortex world volume.


\begin{theacknowledgments}

The authors thank a fruitful discussion and a 
collaboration with Masato Arai, Kazutoshi Ohta, 
Yuji Tachikawa, David Tong, and Yisong Yang. 
This work is supported in part by Grant-in-Aid for Scientific 
Research from the Ministry of Education, Culture, Sports, 
Science and Technology, Japan No.17540237 (N.S.) 
and 16028203 for the priority area ``origin of mass'' (N.S.). 
The works of K.O.~and M.N.~are 
supported by Japan Society for the Promotion 
of Science under the Post-doctoral Research Program  
while the works of M.E.~and Y.I.~are 
supported by Japan Society for the Promotion 
of Science under the Pre-doctoral Research Program. 

\end{theacknowledgments}


\end{document}